\documentclass[twocolumn]{elsart1p}

\usepackage{graphicx}
\usepackage{epsfig}

\usepackage{amssymb}
\usepackage{amsmath}
\begin{document}

\begin{frontmatter}

% Title, authors and addresses

% use the thanksref command within \title, \author or \address for footnotes;
% use the corauthref command within \author for corresponding author footnotes;
% use the ead command for the email address,
% and the form \ead[url] for the home page:
% \title{Title\thanksref{label1}}
% \thanks[label1]{}
% \author{Name\corauthref{cor1}\thanksref{label2}}
% \ead{email address}
% \ead[url]{home page}
% \thanks[label2]{}
% \corauth[cor1]{}
% \address{Address\thanksref{label3}}
% \thanks[label3]{}

\title{Lattice with a Twist : \\
Helical Waveguides for Ultracold Matter}

% use optional labels to link authors explicitly to addresses:
% \author[label1,label2]{}
% \address[label1]{}
% \address[label2]{}

\author{M. Bhattacharya}

\address{Department of Physics, University of Arizona, Tucson, Arizona 85721}

\ead{mb399@email.arizona.edu}

\begin{abstract}
We investigate the waveguiding properties of the optical interference 
pattern of two counter-propagating Laguerre-Gaussian beams. The number, 
helicity, radius, pitch, depth and frequencies of transverse 
confinement of the waveguides are simply related to the beam parameters. 
Quantitative connections to the familiar Gaussian optical lattice are made
and an application to quantum transport is suggested.
\end{abstract}

\begin{keyword}
% keywords here, in the form: 
Optical lattices \sep  waveguides \sep ultracold atoms and molecules

% PACS codes here, in the form: \PACS code \sep code
\PACS 03.75.Be \sep 03.75.Lm \sep 84.40.Az \sep 73.21.Cd

\end{keyword}
\end{frontmatter}
\section{Introduction}
Optical lattices arising from the interference of laser 
beams have proved to be a versatile tool for probing the physics of periodic 
systems \cite{jessen1996}. When loaded with cold atoms or molecules they can 
be used to investigate paradigms ranging from single particle Bloch physics 
\cite{jessen1996} to coherent \cite{morsch2006} and strongly correlated 
many-body systems \cite{lewenstein2006}. Compared to their counterparts in 
the solid-state, optical lattices can be modulated easily in time as well 
as space, possess no phonons and are nearly free from defects.

The most easily available laser modes are Gaussian (G), whose phase structure 
is essentially that of plane waves. Their interference therefore gives 
rise to optical lattices with \textit{discrete} symmetries, which have been 
explored quite intensely \cite{jessen1996,morsch2006,lewenstein2006}. Cold 
atoms loaded into these lattices can be used to simulate other physical 
systems which share the same discrete symmetries, such as the Hubbard model 
on a honeycomb lattice \cite{parmekanti2006}, or quasi-crystals 
\cite{grynberg1997}. However there are also interesting physical systems 
which possess \textit{continuous} symmetries, such as a Bose-Einstein 
condensate (BEC) in a toroidal potential expected to display a persistent 
current \cite{kurn2005,juha1998} or a particle in a helical waveguide 
expected to be bound geometrically \cite{exner2007}. In order to simulate 
these scenarios with cold atoms optical lattices with continuous symmetries 
are required. Such lattices can be made by interfering laser modes with 
azimuthal phase structure such as Laguerre-Gaussian (LG) beams which
have recently become available (\cite{nOAMbook} and references therein).

Some work has been done regarding continuous optical lattices (see below); 
however they are yet to be explored in as much detail as discrete lattices. 
In this Communication we examine the structure of the optical lattice arising 
from the interference of two LG beams. While it is generally known that this 
produces a helical potential for ultracold matter, the waveguiding properties 
of this potential have hitherto been unexplored. Here we provide an elementary 
demonstration of the fact that the lattice can indeed work as a waveguide. 
Further we show that the confinement in both transverse directions in the 
continuous helical lattice can be related to that in the more familiar 
one-dimensional discrete optical lattice via simple geometrical factors. 
Lastly, we present a compact expression for the aspect ratio of the waveguide. 
This enables us to prove that the waveguide is always anisotropic. It also 
provides a convenient but realistic limit to spectral geometers analysing the 
potential for its ability to support geometrically bound states \cite{exner2007}.

To place our work in perspective we point out that theoretically the 
superposition of LG beams has been considered in the context of discrete 
circular lattices \cite{nienhuis2004}, ring lattices without \cite{arnold2006} 
and with \cite{amico2005} tunable boundary phase twists and twisted optical molasses 
\cite{carter2006}. Toroidal traps \cite{ndholakia2001} and atom waveguides 
\cite{dholakia2000} have been proposed using LG beams and also seem achievable 
with the use of microfabricated optical elements \cite{ertmer2001}. 
Experimentally LG modes have been used for waveguiding a BEC \cite{sengstock2001}
and their interference has been used to trap microparticles \cite{macdonald2002} 
as well as to create vortex states in a BEC \cite{Phillips2007,staliunas2002}.

\section{Interference of Laguerre-Gaussian beams}

We consider the intensity pattern arising from the interference of two identical 
linearly polarized LG beams with charge $l$ and a single radial node at the origin, 
counter-propagating along the cylindrical $z-$axis
\begin{equation}
\label{eq:inet} I=4I_{l}(\rho)\cos^{2}\left(kz-l\phi
\right)\cos^{2}(\omega t),
\end{equation}
where $I_{l}(\rho)$ is the intensity due to either beam, $k$ the wavevector 
and $\omega$ the optical frequency. In writing Eq.(\ref{eq:inet}) we have 
considered displacements much smaller than the Rayleigh range, i.e. 
$z/z_{R} \ll 1$, where $z_{R}=\pi \omega_{0}^{2}/\lambda$, $\omega_{0}$ being 
the beam waist and $\lambda$ the wavelength of light.

\section{The helical dipole potential}
We now examine what happens when a cold atom is brought into a region 
illuminated by light with the intensity distribution (\ref{eq:inet}). Without 
loss of generality we consider an isolated ground-excited transition for the 
atom. We let the frequency $\omega$ of the LG beams be red-detuned 
($\Delta < 0$) by $\Delta\gg \Gamma$ from the atomic transition, where 
$\Gamma$ is the linewidth of the excited state. In this case the cold atom 
experiences a nondissipative attractive potential \cite{meystrebook}
\begin{equation}
\label{eq:potential}U=U_{l}(\rho)\cos^{2}\left(kz-l\phi\right),
\end{equation}
where $U_{l}(\rho)=\frac{\hbar
\Gamma^{2}}{2\Delta}\left[\frac{I_{l}(\rho)}{I_{Sat}}\right] < 0$,
and $I_{Sat}$ is the saturation intensity for the transition. The 
non-resonant potential $U$ attracts atoms to the most intense part of its 
intensity distribution. $U$ can trap sufficiently cold molecules as well.

\subsection{The geometry of the potential}
\begin{figure}
\includegraphics*[width=0.49 \textwidth]{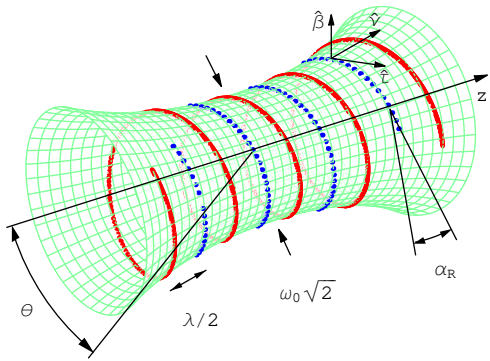}
\caption{\label{fig:Helix}The double-helical lattice, arising from the 
optical interference of two Laguerre-Gaussian $l=1$ beams. 
Two right-handed helical waveguides drawn using solid red and dotted 
blue curves are formed, each with radius proportional to the beam 
radius $\omega_{0}$ and with pitch equal to the optical wavelength 
$\lambda$. They are translated by $\lambda/2$ and rotated by $\pi$ 
with respect to each other. The half-angle of divergence of either 
beam $\theta$, and the pitch angle or \textit{chirality} of each 
helical strand $\alpha_{R}$ are depicted in the figure. Also shown 
are the orthogonal unit vectors of the right-handed Frenet frame 
local to the helix, consisting of the tangent $(\hat{\tau})$, 
normal $(\hat{\nu})$ and binormal $(\hat{\beta})$ at a point.}
\end{figure}

The minima of the potential (\ref{eq:potential}) lie on $2l$ intertwined 
helices of radius \cite{macdonald2002}
\begin{equation}
\label{eq:radius}
R=\omega_{0}\sqrt{l/2},
\end{equation}
and pitch 
\begin{equation}
\label{eq:pitchwithr}
 P=l\lambda\left[1-\left(\frac{3l+2}{4}\right)\theta^{2}\right]^{-1}.
\end{equation}
The pitch of each helix is practically $\sim l\lambda$ ; the correction to it has been 
calculated by including terms linear in $z/z_{R}$ and has been defined 
in terms of the asymptotic half-angle of divergence of either LG beam
\cite{siegmanbook}
\begin{equation}
\label{eq:halfangle} \theta = \frac{\lambda}{\pi w_{0}},
\end{equation}
as shown in Fig.\ref{fig:Helix} which displays the potential for the case
$l=1$. For regular laser beams $\theta$ is quite small, about $10^{-3}$. 
We note that $\theta = 1$ corresponds to the diffraction limit 
\cite{siegmanbook}. 

The helices are displaced $\lambda/2$ parallel to each other along the 
$z$-axis, and rotated $\pi/l$ radians away from each other on any plane 
where $z$ is constant. For high enough laser power because of the red 
detuning $(\Delta < 0)$ a cold atom moves along a particular minimum of 
the potential Eq.(\ref{eq:potential}) and effectively sees a potential 
that is helical in symmetry.

\subsection{Atomic motion along the waveguide}
Here we establish the fact that $U$ can act as a waveguide. Without loss of 
generality we restrict ourselves to a single strand of the ensemble. For 
an atom in such a helix an orthogonal co-moving coordinate system is 
supplied by the Frenet frame consisting of the unit vectors
$(\hat{\tau},\hat{\nu},\hat{\beta})$, which are the tangent, normal and 
binormal to the helix respectively at a point 
$(\rho,\phi,z)$ (Fig.\ref{fig:Helix}) \cite{aminovbook}
\begin{eqnarray}
\label{eq:Frenetframe}
\begin{array}{lll}
\hat{\tau}&=&\sin \alpha
\hspace{0.02in} \hat{z}+\cos \alpha \hspace{0.02in} \hat{\phi}\\
\hat{\nu}&=&\cos \alpha
\hspace{0.02in} \hat{z}-\sin \alpha \hspace{0.02in} \hat{\phi}\\
\hat{\beta}&=& \hat{\rho}, \\
\end{array}
\end{eqnarray}
where the transformation from the cylindrical unit vectors involves the
pitch angle $\alpha$ of a helix of radius $\rho$ defined by
\begin{equation}
\label{eq:pitchangle} \tan \alpha = \frac{P}{2\pi \rho},
\hspace{.1in} 0 \leq \alpha < \pi/2.
\end{equation} 
If the laser detuning ($\Delta$) from the atomic transition is large 
enough, photon scattering due to spontaneous emission is negligible 
and the potential Eq.(\ref{eq:potential}) is essentially conservative. This 
implies that we can calculate the force due to the potential simply as
\begin{equation}
\label{eq:force}
F=- \nabla U.
\end{equation}
Expressing Eq.(\ref{eq:force}) in the Frenet frame Eq.(\ref{eq:Frenetframe}) 
we find that the component of force along the $\hat{\tau}$ direction 
vanishes identically. However there are constraining forces in the
directions $\hat{\nu}$ and $\hat{\beta}$. The potential $U$ 
therefore acts as a waveguide that confines the atom in the locally 
transverse directions while allowing free motion along the local 
longitudinal direction. It is worth noting that this is true 
everywhere in the potential, and not only at its minimum.

\subsection{Atomic motion transverse to the waveguide}
We now turn to a detailed consideration of the atomic motion in the transverse 
directions $\hat{\nu}$ and $\hat{\beta}$. From Eq.(\ref{eq:pitchangle}) 
we find that at the minimum $(\rho=R)$ of the potential $U$ the pitch angle, 
or the \emph{chirality} of the helix (Fig.\ref{fig:Helix}), 
is $\alpha_{R}$ where
\begin{equation}
\label{eq:alphaR}
 \tan \alpha_{R} = \frac{P}{2\pi R} = \theta \sqrt{\frac{l}{2}}.
\end{equation}
To investigate the nature of the potential seen in the transverse 
direction by the atom, we use Eq.(\ref{eq:alphaR}) and the curvilinear 
coordinates local to the Frenet frame
\begin{eqnarray}
\label{eq:ccords}
\begin{array}{ll}
\nu=&(kz-l\phi)/k, \\
\beta=&\rho-R. \\
\end{array}
\end{eqnarray}
Using these quantities we Taylor expand Eq.(\ref{eq:potential}) near its 
minimum (where $\nu \sim 0, \beta \sim 0$, and $\alpha \sim \alpha_{R}$) and obtain 
\begin{equation}
\label{eq:appu}
U \sim U_{l0}(R)+\frac{1}{2}m\omega_{\nu}^{2}\nu^{2}+\frac{1}{2}m\omega_{\beta}^{2}\beta^{2}.
\end{equation}
Thus to lowest order in $\nu$ and $\beta$ the transverse confinement is simple 
harmonic. Let us consider the trapping frequencies $\omega_{\nu}$ and 
$\omega_{\beta}$. The trapping frequency in the $\hat{\nu}$ direction is 
given by 
\begin{equation}
\label{eq:onu}
\omega_{\nu}=\frac{1}{\hbar}\left[ 4 E_{R} U_{l0}(R)\sec \alpha_{R}\right]^{1/2},
\end{equation}
where we have used the single photon recoil energy
\begin{equation}
\label{eq:Erec}
 E_{R}=\frac{\hbar^{2}k^{2}}{2m}
\end{equation}
to write $\omega_{\nu}$ in a transparent form. In Eq.(\ref{eq:Erec}) $m$ is 
the atomic mass. Importantly, Eq.(\ref{eq:onu}) allows us to relate the helical 
lattice to the more familiar one-dimensional optical lattice if we recall that $l=0$ 
corresponds to the interference of G beams. If we put $l=0$ in Eq.(\ref{eq:alphaR}), 
the tangent vanishes. This implies $\sec \alpha_{R}=1$, and Eq.(\ref{eq:onu}) 
reduces to 
\begin{equation}
\label{eq:onu1d}
\omega^{1d}=\frac{1}{\hbar}\left[ 4 E_{R} U_{00}(R)\right]^{1/2},
\end{equation}
which is the trapping frequency for an atom in the well of a 1-d lattice 
constructed from G beams. In this case the $\hat{\nu}$ direction points 
along the $z-$axis. Now for LG beams, $l \neq 0$. Nevertheless since 
$\theta \sim 10^{-3}$ is usually a small quantity, the tangent in 
Eq.(\ref{eq:alphaR}) is still close to zero and hence $\sec \alpha_{R}$ is 
not much greater than 1. Therefore we may not expect the trapping 
frequency in the normal direction for a helical lattice to be very 
different than that for the one-dimensional G lattice for 
comparable laser parameters ($U_{00} \sim U_{l0}$). 

The trapping frequency in the $\hat{\beta}$ direction can be written as
\begin{equation}
\label{eq:obeta}
\omega_{\beta}=\theta \frac{\left[ 4 E_{R} U_{l0}(R)\right]^{1/2}}{\hbar} \sim \theta \omega^{1d}.
\end{equation}
Eq.(\ref{eq:obeta}) shows that for $l \neq 0$ and comparable parameters 
($U_{00} \sim U_{l0}$) the confinement along the binormal in the helical 
waveguide is typically weaker than in the corresponding one-dimensional G 
lattice by a factor of $\theta \sim 10^{-3}$.

Relating the continuous helical LG lattice to the discrete G lattice
allows us to transfer some of the well-established physical intuition from
the latter to the former. For example the sub-wavelength confinement of 
atoms, characteristic of deep discrete optical lattices \cite{meystrebook} 
can typically be obtained only along the $\hat{\nu}$ direction in a helical 
lattice.

The ratio of the two transverse frequencies Eqs.(\ref{eq:onu}) and (\ref{eq:obeta}), i.e. 
the aspect ratio $A$ of the waveguide, is given by
\begin{equation}
\label{eq:anisotropy} A = \frac{\omega_{\nu}}{\omega_{\beta}} =
\frac{1}{\theta} \left(1+\frac{l\theta^{2}}{2}\right)^{1/4}.
\end{equation}
Eq.(\ref{eq:anisotropy}) enables us to examine the cross section of the helical
waveguide quite generally. For example perfect isotropy ($A=1$) requires 
$\theta = 1$ and $l=0$. The first condition corresponds to the diffraction 
limit and the second to the absence of any helical structure. This implies 
that perfect isotropy for the waveguide is not possible even in principle. 
Staying in the diffraction limit and allowing for just a single pair 
of helices ($l=1$) introduces a $10\%$ anisotropy: $A=1.1$. For typical 
experimental parameters however $\omega_{\nu} \sim 10$ MHz, 
$\omega_{\beta} \sim 10$ KHz and $A \sim 10^{3}$, i.e. the anisotropy is quite 
large. An important implication of this anisotropy is the ability of 
the waveguide to serve as a three, two or one-dimensional structure for 
microscopic objects. For sufficiently cold atoms or molecules 
($T < \hbar \omega_{\nu}/k_{B} \sim 10\mu$K) the motion along the $\hat{\nu}$ 
direction can be frozen out and the waveguide is rendered essentially 
two dimensional. Similarly, for even lower temperatures 
($T < \hbar \omega_{\beta}/k_{B} \sim 10 $nK) the atomic motion along the 
$\hat{\beta}$ direction can be frozen out and the waveguide made 
one-dimensional. Here $k_{B}$ is Boltzmann's constant.

\section{Applications}

The helical potential (\ref{eq:potential}) allows access to a less-explored 
symmetry, and may reveal interesting physics such as negative group 
velocities \cite{calvo1978} and Berry's phase \cite{nbitter1987} for particles
moving under its influence. Below we outline a possible application to quantum 
transport.

The availability of a waveguide that bends and twists naturally suggests a 
problem that has been considered (as with much of optical lattice physics)
initially in the condensed matter literature. This is the phenomenon of 
\textit{geometrically bound states} relevant to electron transport through 
nanoscopic wires in quantum heterostructures (\cite{londerganbook} and
references therein). The basic idea can be understood by considering the 
transmission of a quantum particle through a straight waveguide of uniform 
cross-section. A low energy cut-off for the transmission exists because the 
transverse wavelength of the particle cannot be larger than the width of the 
waveguide. Now imagine a bulge in the waveguide, which locally allows for a 
longer wavelength, i.e. an energy lower than the cut-off. In an infinitely long 
waveguide this would correspond to the localization of the particle at the bulge, 
or a bound state. In a waveguide of finite extent, this shows up as an exponentially
decaying resonance : the particle eventually leaks out of the waveguide. The 
same effect can be achieved even in a waveguide of constant width, if the 
waveguide is bent \cite{goldstone1992}. In fact it has been shown quite 
generally that curvature in a waveguide is equivalent to the presence of 
an attractive potential \cite{exner1989}. Classically analogous 
`trapped' electromagnetic modes have experimentally been observed in bent 
microwave waveguides \cite{murdock1992}.

Motivated by our work, Exner \textit{et al.} have examined a curved helical 
potential for the presence of bound states and suggested an answer in the 
affirmative \cite{exner2007}. Their analysis implies that for 
large aspect ratios $(A\gg 1)$ and small pitch angles ($\alpha_{R} \sim 0)$ 
appropriate to the optical potential $U$ described here, particles can be 
bound by the waveguide if the radius goes through a local maximum. It 
would be interesting to test this prediction as in practice the pitch angle of $U$ 
can be adjusted using $\theta$ and $l$ while local extrema in the radius can 
easily be obtained using (de)focussing lenses. One experimental signature of the 
bound state would be a minimum in the transmission through the waveguide at 
the binding energy \cite{londerganbook}.

\section{Conclusion}
We have investigated the waveguiding properties of the optical lattice 
arising from the interference of two identical counter-propagating 
Laguerre-Gaussian beams. We have provided a quantitative connection to the more 
familiar Gaussian lattices. We have also provided a basic description of atomic 
motion in the waveguide.
\section{Acknowledgements}
It is a pleasure to thank P. Meystre for both intellectual and material 
support and E. Wright for stimulating discussions. This work is supported 
in part by the US Army Research Office, NASA, the National Science 
Foundation and the US Office of Naval Research.
% The Appendices part is started with the command \appendix;
% appendix sections are then done as normal sections
% \appendix

% \section{}
% \label{}

\end{document}